\documentstyle[PASJadd,epsf]{PASJ95}
\newcommand{\Mdot}{{\dot M}}

\begin{document}
\setcounter{page}{1}

\title{Theoretical Models of Multi-waveband QSO Luminosity Functions}

\author{Takashi {\sc Hosokawa}$^1$,
 Shin {\sc Mineshige}$^2$, Toshihiro {\sc Kawaguchi}$^1$,
Kohji {\sc Yoshikawa}$^1$, and Masayuki {\sc Umemura$^3$}
\\
{$^1$\it Department of Astronomy, Graduate School of Science, Kyoto
         University, Kitashirakawa, Sakyo-ku, Kyoto 606-8502}
\\
{\it  E-mail(TH): hosokawa@kusastro.kyoto-u.ac.jp}
\\
{$^2$\it Yukawa Institute for Theoretical Physics, Kyoto
         University, Kitashirakawa, Sakyo-ku, Kyoto 606-8502}
\\
{$^3$\it Center for Computational Physics, University of Tsukuba,
1-1-1 Tennoudai, Tsukuba, Ibaraki 305-8577}
}

\abst{Cosmological evolution of the QSO luminosity functions (LFs) 
xat NIR/optical/X-ray bands for $1.3 \ltsim z \ltsim 3.5$ is 
investigated based on the realistic QSO spectra.
The accretion-disk theory predicts that 
although QSO luminosities only depend on mass-accretion rate, $\Mdot$,
QSO spectra have a dependence on black-hole mass, $M_{\rm BH}$, as well.
The smaller $M_{\rm BH}$ is and/or the larger $\Mdot$ is,
the harder becomes the QSO NIR/optical/UV spectrum.  We model
disk spectra which can reproduce these features and 
calculated LFs for redshift $z \sim 3$ 
with the assumption of new-born QSOs being shining at
the Eddington luminosity.  The main results are:
(i) the observed LFs at optical and X-rays can be simultaneously reproduced.
(ii) LFs at optical and X-ray bands are not sensitive to $M_{\rm BH}$,
while LFs at NIR bands are; 
about one order of magnitude difference is expected in volume number densities
at $L_{I, J} \sim 10^{46} \rm{erg \, s^{-1}}$ between the case that
all QSOs would have the same spectral shape as 
that of $M_{\rm BH} = 10^{9} M_{\odot}$ and 
the case with $M_{\rm BH} = 10^{11} M_{\odot}$.
(iii) The resultant LFs at NIR are dominated by $10^{7} M_{\odot}$ 
black-holes at $L_{I, J} \ltsim 10^{44} \rm{erg \, s^{-1}}$, 
and by $10^{11} M_{\odot}$ black-holes 
at $L_{I, J} \gtsim 10^{46} \rm{erg \, s^{-1}}$.
Future infrared observations from space(e.g.NGST) will probe 
cosmological evolution of black hole masses.
For redshift $z < 3$, on the other hand,
the observed optical/X-ray LFs can be fitted, if the initial
QSO luminosity $L_0$ is below the Eddington luminosity $L_{\rm Edd}$.
Interestingly, the best fitting values of $\ell \equiv L_0/L_{\rm Edd}$ 
are different in B- and X-ray bands;
$\ell_{\rm B} \approx 2.5 \, \ell_{\rm x}$. 
The reason for this discrepancy is briefly discussed.
%The reason for this discrepancy is not yet clear, and it would require
%further improvements in observations and/or spectral models.
}

\kword{accretion, accretion disks --- black holes ---
cosmology: theory --- galaxies: active --- galaxies: Seyfert}

\maketitle

\section{Introduction}

The cosmological evolution of QSOs and QSO black holes
attracts many researchers recently
thanks to the rapid progress in observational studies of distant QSOs,
which promotes intensive discussion on
their formation mechanism and formation epochs 
(e.g., Rees 1984; Haiman \& Loeb 1999: hearafter HL99; 
Kauffmann \& Haehnelt 2000; Nulson \& Fabian 2000).
The efforts have been concentrated on understanding the
specific evolutionary behavior of the bright QSO population. 
It shows a rapid increase from the present day back to
redshift $z \sim 2.5$ (Schmidt, Green 1983).
It is still controversial whether the comoving density of
QSOs remained high at $z > 2.5$, as is suggested by $ROSAT$ X-ray
studies (Miyaji et al. 1998, 2000) or rapidly decreases toward higher $z$,
as is obtained by optical and radio survey(Pei 1995,
see also Osmer 1982; Shaver et al. 1996,1999).
This apparent discrepancy may simply imply that
X-ray surveys could detect optically faint QSOs
whose population shows no rapid decline 
beyond $z \sim 2.5$ (HL99).
It is interesting to note similar evolutionary behavior
of the cosmic star-formation history 
(Madau et al. 1996; Connolly et al. 1997; Glazebrook et al. 1999;
Steidel et al. 1999), 
indicating a close link between the QSO evolution
and the cosmic star formation history (Franceschini et al. 1999).

LFs of bright QSOs are well approximated
by double power laws (e.g. Boyle et al. 1988).  This shape contrasts
those of normal galaxies which show exponential decline at high luminosities.
Pioneered by Cavaliere \& Szalay (1986) and Efstathiou \& Rees (1988),
theoretical modeling of the QSO LFs has been 
attempted by several authors; e.g.,
Haehnelt et al. (1998, hereafter HNR98), Haiman \& Loeb (1998,
hereafter HL98; HL99), Cavaliere \& Vittorini (1998, 2000), and
Kauffman \& Haehnelt (2000).  
We specifically pick up models by HNR98 and HL98, who consider
formation of black holes in hierarchically growing dark halos,
whose formation rate is assumed to be described by the
Press-Schechter theory.
It is also assumed that each dark halo
finally produces one supermassive black hole
with mass being a function of mass of its host dark halo, and that
QSO black holes shine at Eddington luminosity for certain period 
by accreting environmental gas 
and then quickly fade out as accretion material is depleted.
The basic shapes of the LFs in either B-bands or X-rays are then
nicely reproduced. 

 However, these studies assume rather ad hoc accretion spectra
which require improvement by using realistic accretion flow models.
This is the motivation of the present study.
We, here, especially focus our discussion on
how to extract information regarding the cosmological evolution 
of black-hole mass from multi-wavelength LFs.

The basic accretion theory tells that
black-hole accretion produces energy output of the order of
$0.1 {\dot M}c^2$ (with $c$ being speed of light); that is,
the disk luminosity, $L$, does not contain information 
regarding black-hole mass except for
the lower limit of $M_{\rm BH}$ set by the constraint that
$L$ should not largely exceed the Eddington luminosity,
$L_{\rm Edd}= 1.5 \times 10^{38} (M_{\rm BH}/M_{\odot})$erg~s$^{-1}$.
We wish to emphasize, however, that
spectral energy distribution certainly depends on black-hole mass.
This is observationally clear, since galactic black-hole candidates with
$M_{\rm BH} \sim 10 M_\odot$ exhibit distinct spectra 
in optical to soft X-ray ranges from those of typical AGNs.  
Further, there is growing evidence that
narrow-line Seyfert 1 galaxies (NLS1s), which are believed to 
harbor relatively less massive black holes 
with $M_{\rm BH} \simeq 10^5 - 10^7 M_\odot$,
show unique soft X-ray features; i.e., enhanced
soft X-ray excess and large spectral index in X-rays
(Pounds et al. 1996; Otani et al. 1996; Boller et al. 1996;
Brandt et al. 1997).

These features can be basically accounted for in terms of the
standard-disk theory (Shakura \& Sunyaev 1973)
and that with some extension (Mineshige et al. 2000).  
It is well known that 
the effective (or surface) temperature of optically thick (standard) disks
depends on $M_{\rm BH}$ as
\begin{equation}
\label{Teff}
 T_{\rm eff}(r) \propto \frac{M_{\rm BH}^{1/4}\dot{M}^{1/4}}{r^{3/4}}
   \propto M_{\rm BH}^{-1/4}
          \frac{(L/L_{\rm Edd})^{1/4}}
               {(r/R_{\rm g})^{3/4}},
%          \left(\frac{L}{L_{\rm Edd}}\right)^{1/4}
%          \left(\frac{r}{R_{\rm g}}\right)^{-3/4},
\end{equation}
that is, the peak frequency of the disk spectral peak shifts
in proportion to $M_{\rm BH}^{-1/4}\times (L/L_{\rm Edd})^{1/4}$.
The smaller $M_{\rm BH}$ is and/or the larger $\Mdot$ (or $L$) is,
the harder becomes the QSO optical/UV spectrum.  
X-ray spectra, in contrast, practically have no mass dependence,
since the electron temperature of optically thin, hot plasmas near black holes
is in any cases kept around $5\times 10^9$ K regardless of black hole mass.
This is explicitly demonstrated in the framework of a model of 
optically thin, advection-dominated accretion flow (ADAF; Ichimaru 1977,
see Manmoto et al. 1997 for the spectrum without self-similar assumptions). 

Then, we may be able to extract information concerning black-hole masses
as a function of redshift, $z$,
through the comparisons of QSO LFs
at two (or more) different wavebands; 
e.g. X-rays, optical, and infrared bands.
This will be a final goal of the present study.
The plan of the present paper is as follows:
in section 2 we explain our models, which are basically the same as those
of HL98 and HNR98 except for adopting realistic QSO spectral models.
We then give our resultant LFs at B-, X-ray, and NIR bands
for $z \sim 3$ and $z \lsim 2$ in sections 3 and 4, respectively.
The final section is devoted to discussion.

\section{Our Models}

\subsection{Basic assumptions}

It is generally believed that most of the mass in the universe
is in the form of ``dark matter (DM).''
Accordingly,
the spatial distribution of luminous objects, such as 
galaxies, QSOs, and so on, is likely to
follow that of DM.  Therefore, it is reasonable
% to use the number of dark halos per unit comoving volume
to use the volume number density of dark halos 
to investigate that of luminous objects, including QSOs.
We here adopt two simple models for calculating LFs,
following HNR98 and HL98.
The basic assumptions made in both models are as follows:

\subsubsection{Relation between halo mass and black-hole mass}
We assume that each dark halo 
necessarily has only one supermassive black hole at the center,
though it is still controversial observationally.
Then, we have
\begin{equation}
M_{\rm BH} = F(M_{\rm halo})  
\end{equation}
where 
%$M_{\rm BH}$ is the black-hole mass,
$M_{\rm halo}$ is the host dark-halo mass
and the function form $F(M)$ depends on models, as prescribed below:

In model A (HNR98),
the relation between $M_{\rm BH}$ and $M_{\rm halo}$ is nonlinear; 
\begin{equation}
\label{modelA}
M_{\rm BH} =C  v_{\rm halo}^5 = C'  M_{\rm halo}^{5/3} (1+z)^{5/2},
\end{equation}
where $C'$ and $C$ are constants and $C$ is treated as a free parameter. 
The physical meaning of this relation is described in Silk \& Rees (1998).
Equation (\ref{modelA}) gives a critical black hole mass 
to bound gas in its host halo against the outward wind from QSO. 
If black hole mass exceeds this upper limit, 
QSO could expel all the gas from its host galaxy,
and the black hole will never grow further.

In Model B (HL98), in contrast,
a linear relation between $M_{\rm BH}$ and $M_{\rm halo}$ is assumed;
\begin{equation}
\label{modelB}
M_{\rm BH} = \epsilon M_{\rm halo}.
\end{equation}
Here, $\epsilon$ is a free parameter.
This expression is based on the observed linear relation 
among the bulge luminosity and the black-hole mass 
(Magorrian et al. 1998, hereafter MG98; see also Laor 1998;
Ferrarese \& Merritt 2000; Gebhardt et al. 2000).

\subsubsection{QSO light curves}
We, next, assume that
time evolution of the QSO luminosity at high $z$ ($\gsim 3$) follows
\begin{equation}
\label{L}
 L(t) = L_{\rm Edd} \exp \left( - \frac{t}{t_{\rm Q}} \right)
       \equiv M_{\rm BH}~g(t),
\label{lumi}
\end{equation}
where we set $t=0$ when a halo collapses. 
Such a simple prescription for a single
QSO light curve is known to reproduce well
the observed LFs at $z \gsim 3$
(HL98, HNR98).

If one directly uses the models described above to calculate
QSO LF at $z < 2$, 
we over-predict number density of QSOs
compared to the observed co-moving density of QSOs that exhibits
rapid decline towards $z=0$ from $z \sim 2.5$.
Therefore, model assumptions should be re-examined.
The rapid decay in the QSO density seems to be caused by
the depletion of fueling mass; i.e., environmental gas
surrounding a black hole is not enough to shine
the black hole at $L_{\rm Edd}$.
Hence, it is reasonable to assume that
the QSO luminosity should be less than $L_{\rm Edd}$
even at the epoch when its host halo collapses
(Mcleod et al. 1999; Haiman \& Menou 2000).  We thus
% introduce a parameter, $\ell \equiv L_0/L_{\rm Edd} < 1$, and,
introduce a parameter, $\ell \equiv L_0/L_{\rm Edd} \leq 1$, and,
assuming a relation, $L_0 \propto L_{\rm Edd} \propto M_{\rm BH}$, we set
\begin{equation}
 L(t) =  L_0 \exp \left( - \frac{t}{t_{\rm Q}} \right)
      \equiv \ell M_{\rm BH}~g(t). 
\label{lowl}
\end{equation} 
Considering the evolution of gas content in galaxies due to star
formation activity, $\ell$ will be decreasing towards $z \sim 0$.

\subsubsection{Formation rate of dark halo}
In previous studies, the formation rate of dark halos and black holes was 
often regarded as the time derivative of Press-Schechter (PS) mass function
(Press \& Schechter 1974); 
\begin{equation}
\label{PS}
     \frac{d^2N_{\rm BH}}{dM_{\rm BH}dz} = \frac{1}{\epsilon}
     \frac{d}{dz} \frac{dN_{\rm PS}}{dM_{\rm halo}}.
\end{equation}
At later times (especially at $z \lsim 2$), however, this
is inadequate, since equation (\ref{PS})
gives a negative formation rate for small halo masses.
This is because that
the time derivative of PS mass function contain two distinct terms:
i) the formation rate (which is positive)
of halos with mass $M$ from objects with lower masses 
and ii) the destruction rate (which is negative)
due to 
merging of halos with mass $M$ into objects with higher masses, and 
the latter dominates over the former at later times for small halo masses. 
Many authors have attempted to derive the genuine formation rate 
in various ways (e.g., Lacey \& Cole 1993, Sasaki 1994).
We here adopt a model proposed by Kitayama \& Suto (1996).
Then, we regard the formation rate of black holes to be 
\begin{equation}
\label{Form}
 \frac{d^2N_{\rm BH}}{dM_{\rm BH}dz} = \frac{1}{\epsilon}
 \frac{d}{dz} \frac{dN_{\rm form}}{dM_{\rm halo}}
  \times p(z',z). 
\end{equation}
% where, $\frac{d}{dz} \frac{dN_{\rm form}}{dM_{\rm halo}}$
where, $d(dN_{\rm form}/dM_{\rm halo})/dz$
is the formation rate of dark halos, and $p(z',z)$ 
is the {\it survival probability}, probability
that the dark halo formed at $z'$ remains at $z$ without
merging into objects of higher masses.  
Kitayama \& Suto (1996) calculated the genuine formation rate by
the merging rate of halos of $< M/2$ into a halo of mass $M$.
%Here, in this formalism,
%it is noted that ``destruction'' means a halo of mass $M$ merges 
%into a halo of $> 2M$, ``formation'' means a halo of mass $M$ 
%forms from halos of $< M/2$.

After all, the number of free parameters is three for each model: 
QSO lifetime, $t_{\rm Q}$,
the initial Eddington ratio, $\ell$, 
and the constant related to $M_{\rm BH}$,
$C$~(model A) or $\epsilon$~(model B).  
These are to be determined
such that the model should reproduce the observational data most
successfully. 
  
\subsection{QSO Luminosity Functions}
Here, we let $\Phi(L,z)dL$ be the number of QSOs per unit
comoving volume at redshift $z$, whose absolute bolometric luminosity 
is between $L$ and $L+dL$.
On the basis of these assumptions described in
subsection 2.1.1--2.1.3, QSO LFs,
$\Phi(L,z)dL$, can be calculated by the summation of luminosities of
all the QSOs whose luminosity is $L$ at redshift $z$; that is
\begin{eqnarray}
\label{LF}
 \Phi(L,z) \!\!\!&=&\!\!\! 
               \int^\infty_0 \int^z dM_{\rm BH} dz'
               \frac{d^2N_{\rm BH}}{dM_{\rm BH}dz'}    \nonumber\\
        \!\!\!&\times& \!\!\!
	        \delta\left[M_{\rm BH}-\frac{L}{\ell g(t_{z,z'})}\right],
%               &\times& \delta[ L - M_{\rm BH}g(t_{z,z'}) ] ,  
\end{eqnarray}
where $\delta$ is delta-function,
$g(t)$ is defined in equation (\ref{L}), and
$t_{z,z'}$ is the time between the epochs of redshifts $z$ and $z'$.
Here, the lower limit of $M_{\rm halo}$ is 
$M_{\rm halo} \geq 10^8 M_{\odot} [~(1 + z)/10~]^{-1.5}$.
This corresponds to the virial temperature, $T_{\rm vir} \geq 10^4 \rm{K}$.
This is the condition for the gas to sink to the black hole by cooling effect.

Practically, it is convenient to define 
%$f_{\rm band}=L_{\rm band}/L$, 
$L_{\rm band}$, the observational band luminosity.
%and $L$ is the bolometric luminosity, respectively.
We should note that $L_{\rm band}$ is a function of
$M_{\rm BH}$ and $\dot M$ in accordance with mass and mass-flow rate
dependences of the QSO spectra.
We, hereafter, use $\Phi(L_{\rm band},z)$
rather than $\Phi(L,z)$ and simply write
$\Phi(L_{\rm band},z)$ as $\Phi(L,z)$ below.

The LFs certainly depend on the adopted cosmological model.
Though it still remains an open question 
whether the universe is open or closed,
recent observations seem to indicate a flat universe 
(e.g., de Bernardis et al. 2000). 
Thus, here we adopt a $\Lambda$CDM cosmology with tilted power spectrum,
 $(\Omega_0, \Omega_\Lambda, \Omega_{\rm b}, h, \sigma_{8h^{-1}}, n)
= (0.35, 0.65, 0.04, 0.65, 0.87, 0.96)$, as is suggested by
of Ostriker \& Steinhardt (1995).  
%Here, the cosmological parameters
%are mean density in a unit of critical density, $\Omega_0$,
%cosmological constant, $\Omega_\Lambda$, baryon density, $\Omega_{\rm b}$,
%Hubble parameter, $h$, ....., the primordial power spectral index, $n$. 
 
\subsection{Adopted QSO Spectra}

%%% Fig. 1 %%%%%%%%%%
\begin{figure}[t]
 \epsfxsize\columnwidth \epsfbox{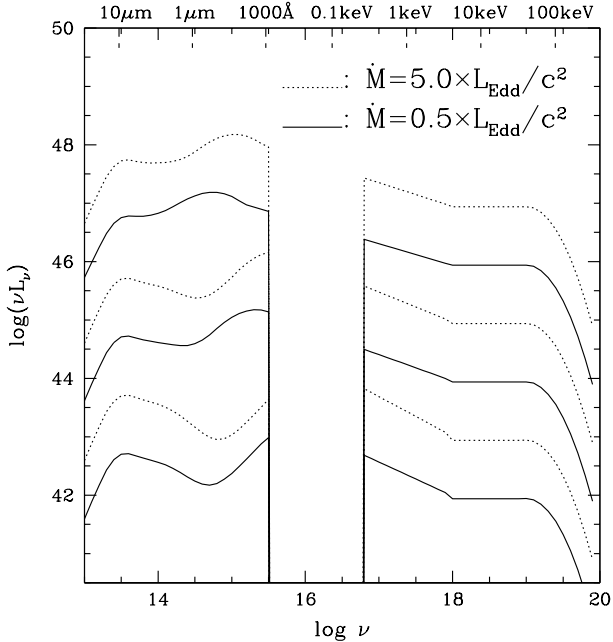}
\caption{The adopted QSO spectra.
The solid lines are the cases for $\dot{M}=0.5 \times  L_{\rm Edd}/c^2$,
while the dotted lines are for $\dot{M}=5.0 \times L_{\rm Edd}/c^2$.  
Three lines of each set represent the spectra for
$M_{\rm BH}=10^{11}, 10^9, 10^7 M_{\odot}$ from the top downwards.  
}
\label{spec}
\end{figure}
%%%%%%%%%%%%%%%%%%%%%%%

As discussed before, we need to give disk spectra to calculate 
band luminosities, $L_{\rm band}$, and thus QSO LFs.
Here, we adopt a simple but reasonably realistic model of 
QSO spectra as displayed in figure \ref{spec} 
These spectra are the sum of three components: 
emission from a disk-corona structure (at $\lambda < 1 \mu$m
or $\log \nu > 14.5$),
that from an outer optically thick disk (around $\lambda \sim 1 \mu$m),
and the IR bump (at $\lambda > 1 \mu$m).  In addition,
we simply assume strong photo-electric absorption due to IGM 
(Especially, Lyman limit systems mainly contributes to absorption
here. (Stengler-Larrea et al. 1995))
in the range, 60 \AA $\leq \lambda \leq$ 912 \AA.

\subsubsection{Disk-corona component}
The disk-corona model spectrum is the main component 
at $\lambda \leq 1\mu$m.
Here, we use a model by Kawaguchi et al. (2000),
since they could, for the first time,
%reproduce the observed broad-band spectral properties of QSOs
%as represented by the composite spectrum 
%(Zheng et al.\ 1997; Laor et al.\ 1997).
reproduce the observed broad-band (optical to hard X-ray) 
spectral energy distributions of QSOs.
The idea is to couple a standard-type disk body with
$T_{\rm eff} \sim 10^5 (r/r_{\rm g})^{-3/4}$ K and
an advection-dominated disk corona, in which $T_{\rm elec} \sim 10^9$ K.
Following Haardt \& Maraschi (1991), they assume that
a fraction, $1-f$, of energy is dissipated in the disk body,
while the remaining fraction, $f$, is in the coronae.  
By solving the hydrostatic balance and 1D radiative transfer
including inverse-Compton processes for each radius,
they finally obtain the spectrum at $4 - 300 r_{\rm g}$.  
According to this model,
the big blue bump (BBB) is by thermal black body emission from the disk 
body at small radii,the soft X-ray excess is inverse-Compton 
scattering of the BBB soft photons in the corona at small radii, 
and the hard X-rays are bremsstrahlung radiation
from the coronae at large radii.

Free parameters involved with this model are
$M_{\rm BH}$, $\dot{M}$, $f$, $\alpha_{\rm c}$, 
$\tau_{\rm c}$, $\tau_0$, $R_{\rm in}$, $R_{\rm out}$; 
black-hole mass, accretion rate, fraction of energy dissipated in the
corona, viscosity parameter in the corona,
coronal optical depth, disk optical depth, inner and outer edges of the corona,
respectively.
The calibration of this model spectra is made so as to
reproduce the so-called QSO composite spectrum
(Zheng et al.\ 1997; Laor et al.\ 1997),
in which the spectral indices ($L_{\nu} \propto \nu^{-\alpha}$) are 
$\alpha \sim 0.3$ at optical (1 $\mu$m $\geq \lambda \geq$ 2500 \AA),
$\alpha \sim 1.0$ at UV(2500 \AA $\geq \lambda \geq$ 1000 \AA),
$\alpha \sim 1.8$ at FUV(1000 \AA $\geq \lambda$),
$\alpha \sim 1.6$ at soft X-ray(0.2--2.0 keV), and
$\alpha \sim 0.7$ at hard X-ray($>$ 2.0 keV), 
and cut-off energy for hard X-ray is about 100 keV.
Kawaguchi et al.\ find that one can
reasonably reproduce the observational QSO composite spectrum
with parameters of
($f$, $\alpha_{\rm c}$, $\tau_c$, $\tau_0$, 
  $R_{\rm in}$, $R_{\rm out}$) =
(0.6, 1.0, 0.6, 1000, 4.0$R_{\rm g}$, 300$R_{\rm g}$),
$M_{\rm BH}=3 \times 10^{9} M_{\odot}$
and $\dot M = 0.5 \times L_{\rm Edd}/c^2$.
Note that $\dot{M}$ of $12 L_{\rm Edd}/c^2$ corresponds to a disk with 
$L_{\rm Edd}$.
The frequency at which the BBB reaches its peak flux
varies as $M_{\rm BH}^{-1/4}$ as in the standard disk model,
since the main component of BBB is
the thermal emission from the optically thick disk
[see equation (\ref{Teff})].

Mass accretion rate is assumed to change according to equation (\ref{lumi}).
Thus, the spectral shape also changes 
with the time elapsed after a halo collapses.
The spectra for different $M_{\rm BH}$ and $\dot M$
are calculated according to the disk corona model with the
same parameter set given above.
As $\dot M$ decreases and/or $M_{\rm BH}$ increases,
the spectral peak shifts towards longer wavelength regimes,
as ${{\dot M}/L_{\rm Edd}}^{-1/4}$ and $M_{\rm BH}^{1/4}$ (Figure 1).
These dependences are the same as that of the standard disk.
Since optical/NIR spectral index $\alpha$ is $\sim 0.3$, luminosity
those bands varies as $M_{\rm BH}^{1.2}$ with fixed ${\dot M}/L_{\rm Edd}$.
The X-ray luminosity is in proportion to $M_{\rm BH}$, on the other hand. 

\subsubsection{Outer optically thick part}
In addition, we consider the spectrum of the standard disk 
at larger radii~($r > 300~R_{\rm g}$), assuming that disk corona
only exists inside $300~R_{\rm g}$. 
This component appears between the IR bump and the BBB in the spectrum. 
For the total disk size of $1000R_{\rm g}$, 
the spectral index of the standard disk is $\alpha \sim -0.3$ 
at this wavelength.

\subsubsection{IR bump}
The second component is the IR bump, typically extending
 from (20--30) to 1 $\mu$m (Telesco et al. 1984; Radovich et al. 1999). 
At the present,
it is widely believed that 
the dust thermal emission is responsible for this bump.
A supporting evidence is that the higher energy end of the 
bump is found at $\sim$ 1 $\mu$m, corresponding to a temperature of
$\sim$ 1800 K, at which dust sublimates (e.g., Kobayashi et al.\ 1993). 
Here, we assume that the spectral indices of
$\alpha = 1.4$ at $\lambda < 10~\mu$m 
(Neugebauer et al. 1987; Polletta \& Courvoisier 1999), and
$\alpha = -2.0$ at $\lambda > 10~\mu$m (Rayleigh-Jeans regime),
and that the total IR power varies with $\dot M$, 
keeping the luminosity ratio of the BBB to the IR bump. 
Although this assumption is rather uncertain,
it is worth noting that this component contributes not to band luminosity,  
but only to bolometric luminosity in the present study.

%%% Fig. 2 %%%%%%%%%%
\begin{figure}[t]
 \epsfxsize\columnwidth \epsfbox{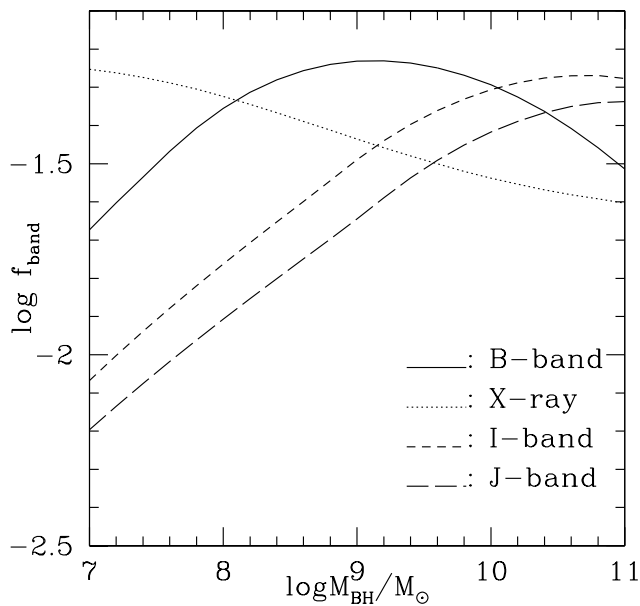}
\caption{
The fraction of the luminosity 
in each band among bolometric luminosity, $f_{\rm band}$, 
as a function of $M_{\rm BH}$ at $z=3.0$ 
and $\dot{M} = 5.0 \times L_{\mbox{Edd}}/c^2$.
Line represent B-band(solid line), X-ray(dotted line), I-band(short 
dash line), and J-band(long dash line), respectively.
}
\label{frac}
\end{figure}
%%%%%%%%%%%%%%%%%%%%%%%

\section{QSO Luminosity Functions at $z \sim 3.0$}

%%% Fig. 3 %%%%%%%%%%
\begin{figure}[t]
 \epsfxsize\columnwidth \epsfbox{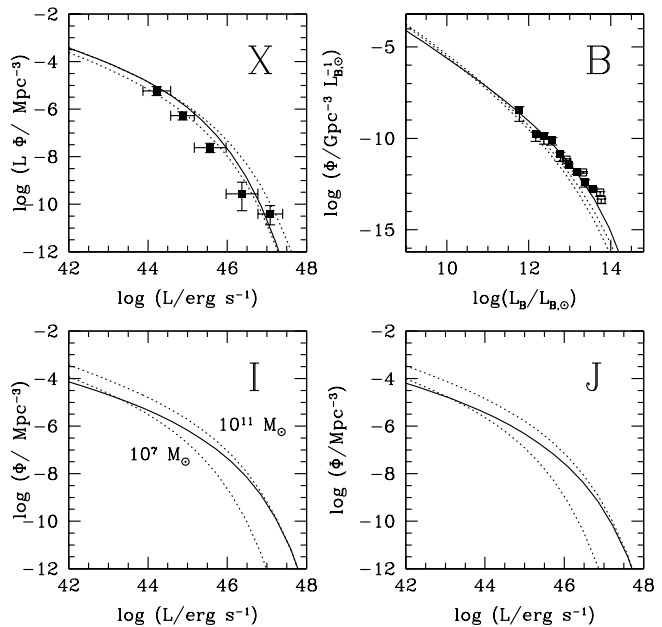}
\caption{QSO Luminosity Functions calculated based on model A. 
Here, model parameters are $t_{\rm Q}=4.0 \times 10^{7}$ yr, 
$\ell$ = 1, and  
$C=4.0 \times 10^8$. 
Note that $z=3.5$ for X-ray $ROSAT$ band, 
and otherwise $z=3.0$. 
Top left; X-ray $ROSAT$ band. 
The filled squares are data by Miyaji et al. (1998) 
for the redshifts of $2.3<z<4.6$.
Top right; B-band. The filled squares are data by Pei (1995) for the
redshifts of $2.5<z<3.5$.
Bottom left; I-band. Bottom right; J-band.
In each figure, dotted lines represent the case that
 QSOs have the same spectrum (and thus the same $f_{\rm band}$)
as that for black hole mass of
$M_{\rm BH} \simeq 10^7$ and $10^{11} M_{\odot}$ at 
$\dot{M} = 5.0 \times L_{\rm Edd}/c^2$~(see Figure 1),
and the solid lines represent the case that each QSO has 
$M_{\rm BH}$-dependent spectrum.}
\label{LFA}
\end{figure}
%%%%%%%%%%%%%%%%%%%%%%%

%%% Fig. 4 %%%%%%%%%%
\begin{figure}[h]
 \epsfxsize\columnwidth \epsfbox{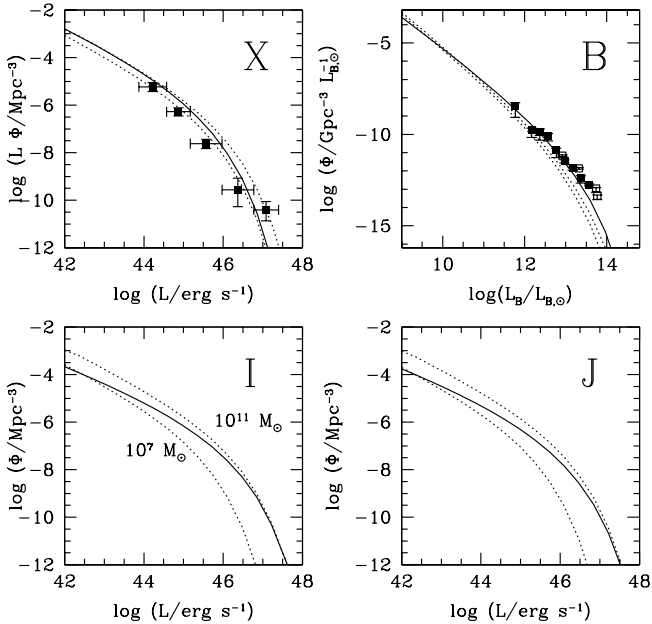}
\caption{Same as figure \ref{LFA} but for model B. 
Model parameters are 
$t_{\rm Q}=3.5 \times 10^{5}$ yr, $\ell$ = 1, 
and $\epsilon=4.0 \times 10^{-4}$.}
\label{LFB}
\end{figure}
%%%%%%%%%%%%%%%%%%%%%%%

Once the spectra are given, we can calculate 
the fraction of the luminosity at a certain band
among the bolometric one, $f_{\rm band}\equiv L_{\rm band}/L$. 
The previous models adopt the same spectral shape taken from
the averaged observational one (e.g., Elvis et al. 1994)
for wide ranges of $M_{\rm BH}$ and $\dot M$;
that is, the value of $f_{\rm band}$ was kept constant.
This treatment is, however, inappropriate, since QSO spectra
should have some $M_{\rm BH}$ and $\dot M$ dependences as we saw
in the previous subsection.
In other words, $f_{\rm band}$ should change as a black hole grows
and $\dot M$ decays with the time.

Figure \ref{frac} summarizes the resultant $f_{\rm band}$ in
B-(4035 -- 4765\AA), I-(0.94 -- 1.12 $\mu$m), J-bands(1.15 -- 1.33 $\mu$m), 
and X-ray~(0.5 -- 2.0 keV, corresponding to the $ROSAT$ energy band).
This figure shows that in the I- and J-bands 
$f_{\rm band}$ decreases by one order of magnitude
as $M_{\rm BH}$ decreases by four orders,
whereas in the B-band and X-ray $f_{\rm band}$ varies less.
This distinct behavior in the infrared bands compared with other bands 
arises from the fact that BBB spectral shape moves towards shorter
wavelength with decreasing $M_{\rm BH}$
Accordingly, this produces interesting features 
in the shape of the LFs.

The theoretical QSO LFs can be calculated 
according to equation (\ref{LF}) for each waveband. 
As noted before, the parameters 
were determined such that the theoretical models should reproduce
the observational, rest-waveband luminosity functions in B-band~(Pei 1995) 
and in X-rays ~(Miyaji et al 1998).
In the present study, we assume that the AGN obscuration play only a minor 
 role in the high-$z$ ($> 1$), high luminosity ($L_{\rm x} > 10^{44.5}$ erg 
 s$^{-1}$) regimes for the following reasons.  It is indeed true that there 
 are several intrinsically obscured QSOs observed (obscured by some material, 
 say, dusty tori and/or starburst regions surrounding the active nuclei). 
 Brandt (1997) noted the existence of type 2 QSOs and Veilleux et al. (1999) 
 pointed out that a part of ULIRG is the candidate for highly obscured QSOs. 
 However, it is unlikely that they contribute significantly to the QSO LFs 
 at the bright end, since Akiyama et al. (2000) reported that type 2 AGNs 
 known to date are dominated by nearby low-luminous objects in 2 -- 7 keV band
 (see their Fig.9).  Further, they plot the redshift distributions of 
 type 1/2 AGNs as functions of redshifts (see their Fig.10b).
 Then, they have calculated the expected redshift distribution of type 1 AGNs 
 using the hard-band X-ray LFs of type 1 AGNs (Boyle et al.1998), finding that
 this expected distribution agrees with that of their samples of type 1 
 AGNs (obtained by ASCA Large Sky Survey) by the possibility of 64\%.
 Next, they did the same but for type 2 AGNs assuming that the shape of
 LFs of type 2 AGNs is the same as that of type 1 AGNs.
 The expected number of the 
 type 2 AGNs at $z>0.4$ is about 10 among 30 identified AGNs in total, 
 whereas there was no type 2 AGNs detection in their data at high redshift.  
 That is, it is very unlikely that type 2 AGNs are distributed in the same way 
 as that of type 1 AGNs; such a probability is only 5\%.  
 Certainly, there exists a deficiency of type 2 AGNs at $z>0.4$,
 at least, in the high luminosity regimes.  Still, we admit that the source 
 number is insufficient to derive solid conclusion from their data.  We should 
 await further observational study to be conducted in near future.

 In the present study, we thus consider only type 1 QSOs in the 
 calculations of the QSO LFs to be compared with the observed QSO LFs.
 It is still possible that, depending on X-ray energy ranges, the fraction of 
 type 2 AGNs may vary to some degree, giving rise to slightly different 
 behavior in LFs at soft and hard X-ray bands.  This issue will be discussed 
 in a future paper.

The observed luminosity functions depend on the cosmological
parameters (Hubble parameter, $h$, and the deceleration parameter, $q_0$),
as well as the spectral index at the relevant bands (for $K$-correction). 
The dependence on $h$ and $q_0$ is simple; 
$\Phi(L,z) \propto dV^{-1} d_{L}^{-2}$,  $L \propto d_{L}^{-2}$ 
(here, $dV$ is the comoving volume element, and $d_{L}$ is the 
luminosity distance).  
As shown by Hartwik \& Schade~(1990), the dependence 
on $\alpha$ is comparatively large. Pei~(1995) presents data
only for $(h,q_0,\alpha)=(0.5,0.5,0.5)$ and $(h,q_0,\alpha)=(0.5,0.1,1.0)$.
In our spectral model, $\alpha$ is nearly 1.0 in the B-band at
redshift, $z \sim 3$.  Thus, we modify Pei's data for 
$(h,q_0,\alpha)=(0.5,0.1,1.0)$ to satisfy $\Lambda$CDM. 
Since the observed optical and X-ray LFs are corrected with an
assumed $\alpha$ of unity, which is equivalent to no K-collection,
we calculate theoretical LFs at observer's frame.
Once the parameters are determined by optical and X-ray LFs, then 
the luminosity function expected in the I- and J-band can be obtained.         

Figure 3 and 4 represent the calculated luminosity functions for model A
and model B, respectively. In all the panels of both figures, 
the dotted lines represent models assuming that 
% all QSOs have the same spectrum as that with black-hole mass of
all QSOs have the same spectrum as that with $M_{\rm BH}$ of
$10^7$ (lower line) or $10^{11}M_{\odot}$~(upper line)
% and mass-flow rate of $\dot{M}=5.0 \times L_{\rm Edd}/c^2$,
and $\dot{M}$ of $5.0 \times L_{\rm Edd}/c^2$,
whereas the solid lines represent models that 
each QSO exhibits different spectra according to variations of 
black-hole mass. 
%
%Note that
%in the X-ray and B-band dotted lines are hardly distinguished
%each other due to different behaviors of $f_{\rm BAND}$ on $M_{\rm BH}$
%(see Figure 2) and $\dot{M}$. 
%That is, LFs in these bands contain practically no
%information regarding black-hole mass, while LFs in infrared bands do.

As figures 3 and 4 show, the dotted lines nearly
coincide with each other in X-ray and B-band, whereas 
they can be clearly separated in I- and J-band.
In other words, LFs in X- and B-bands contain practically no
information regarding black-hole mass, 
while LFs in NIR bands do.
This can be explained in terms of the different 
behavior of $f_{\rm band}$ in those band, as are illustrated 
in figure 2. 
In the X- and B-band, $f_{\rm band}$ varies little 
(within a factor of $\sim$2), 
even if $M_{\rm BH}$ changes by four orders of magnitude,
while in I- and J- bands
$f_{\rm band}$ appreciably decreases (a factor of $\sim$ 10) 
as $M_{\rm BH}$ decrease. 
In the lower two panels of figures 3 and 4, especially,
the solid line approaches the dotted line 
of $M_{\rm BH}=10^{11}M_{\odot}$ on the higher luminosity side 
($L_{\rm band} \gtsim 10^{46}$ erg s$^{-1}$), whereas 
it approaches that of $M_{\rm BH}=10^7M_{\odot}$ 
on the lower luminosity side ($L_{\rm band} \ltsim 10^{44}$ erg s$^{-1}$). 
This clearly points that the more massive the black hole is,
the brighter the QSO becomes at NIR bands (Figure 2).

The dotted lines in figures 3 and 4 are inconsistent with our
model spectra, since those lines are plotted under the assumption
that QSOs with different $M_{\rm BH}$ have the same spectral shape,
while we expect $M_{\rm BH}$-dependence of the QSO spectrum
as is demonstrated in figure \ref{spec}
As mentioned before, the previous studies have used
the one observational spectrum for all QSOs. 
In this sense, the dotted lines also represent the cases, in which
the spectral shape is fixed as in the previous models  (e.g. HL98, HNR98).
The observational mean spectrum by Elvis et al~(1994) is
nearly equal to our spectrum model of 
$M_{\rm BH}\simeq 10^9 M_{\odot}$ and $\dot{M}=5.0\times L_{\rm Edd}/c^2$.
The upper panels of figures 3 and 4
show that whichever model one may use, the previous model or our model, 
the luminosity functions hardly differ in B-band and X-ray. 
The lower panels, on the other hand, demonstrate that
black-hole mass dependence of the QSO spectra manifests in
the luminosity function in I- and J-band. 
Therefore, if QSO luminosity functions are observed
at $z \sim 3$ in these wave bands in future, we can, in principle, constrain 
the $M_{\rm BH}$-dependence of QSO spectrum, thus probing 
cosmological evolution of the black-hole mass in QSOs.
     
Despite the fact that QSO diminishing timescale is longer 
in model A~($t_{\rm Q} \sim 4.0 \times 10^7$ yr) 
than in model B~($3.5 \times 10^5$ yr), 
these figures look very similar. This is because 
the mass fractions are different:
$\epsilon = 4.0 \times 10^{-4}$ in model B,
while in model A $\epsilon$ is not constant but around $\sim10^{-5}$. 
Here, we note that if we use larger $t_Q$ and smaller $\epsilon$ in 
model B, we cannot reproduce observed LFs.
Since the inclination of LFs mainly depends on $\epsilon$,
theoretical LFs become more steeper with smaller $\epsilon$.   

Which value is more reasonable for $\epsilon$?
The ratio of the black-hole mass to the host halo mass ($\epsilon$) 
is not directly observable, but MG98
suggest for the nearby galaxies and QSOs, the ratio of the 
central black-hole mass
to the bulge mass is $M_{\rm BH}/M_{\rm bulge} \sim$ 0.006.     
Then, the value of $\epsilon$ can be estimated through the baryon
mass fraction in the total,
$\Omega_{\rm b}/\Omega_0 \simeq 0.1$. If most of baryonic matter is
contained in the bulge, we have $\epsilon \sim 10^{-3.0}$,
more consistent with the Model B assumption.
MG98 also support the linear $M_{\rm BH}-M_{\rm halo}$
relation, equation (\ref{modelB}).  
On the other hand, Fukugita et al.~(1998) noted that
the bulge density is less than the baryon density by about
one order of magnitude, $\Omega_{\rm bulge}=0.1 \Omega_{\rm b}$. 
If so, we have $\epsilon \sim 6 \times 10^{-5}$,
in good agreement with the Model A assumption. 
More recently,
Meritt \& Ferrarese (2001) noticed the tight $M_{BH}-\sigma$ 
relation (with $\sigma$ being halo velocity dispersion);
$M_{\rm BH} \propto \sigma^{4.72}$, which is close to equation (\ref{modelA}).  
They also show $M_{\rm BH}/M_{\rm bulge} \sim$ 0.0012, 
which is by a factor of $\sim 5$ smaller than the value 
obtained by MG98 and could rule out model B.
The similar conclusion was derived based on the mass estimate
based on the reverbration mapping (Laor 1998).

Another parameter $t_{\rm Q}$ should also constrain models.
By the usual picture of Eddington-limited accretion, 
black hole only grows as $M_{\rm BH} \propto \exp(t/t_{\rm Edd})$, 
where $t_{\rm Edd} \sim 3 \times 10^7$ yr is the {\it e-folding time}.
As HNR98 mentioned, this supports Model A.

%%% Fig. 5 %%%%%%%%%%
\begin{figure}[t]
 \epsfxsize\columnwidth \epsfbox{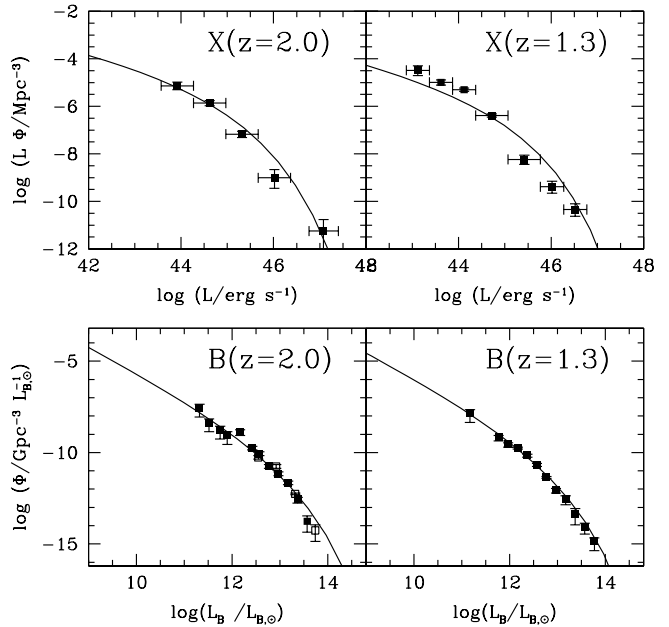}
\caption{
The same as figure 3 (model A) but at different redshifts: 
$z = 2.0$ (solid lines) and $z = 1.3$ (dashed lines), respectively.
and in B-bands (top right). 
Model parameters are
%$t_{\rm Q}= 1 \times 10^6$yr and $\epsilon=2 \times 10^{-4}$, which are
the same as those in figure 4 except for the initial Eddington ratios:
$\ell_{\rm x}=0.24$ ($z=2.0$) and 0.10($z=1.3$), and
$\ell_{\rm B}=0.66$ (left) and 0.25 (right).
Also plotted are the observed LFs.
The filled squares in the top left panel represent 
the observed LFs in X-ray $ROSAT$ bands 
 taken by Miyaji et al.~(1998) at redshift of $1.6 < z < 2.3$ 
and at $0.8 < z< 1.6$, while
the filled and open squares in the top right panel are data
from Pei~(1995) for the redshifts of $1.9 < z < 2.2$ and $1.2 < z < 1.4$. 
}
\label{LowLF}
\end{figure}
%%%%%%%%%%%%%%%%%%%%%%%

%%% Fig. 6 %%%%%%%%%%
\begin{figure}[t]
 \epsfxsize\columnwidth \epsfbox{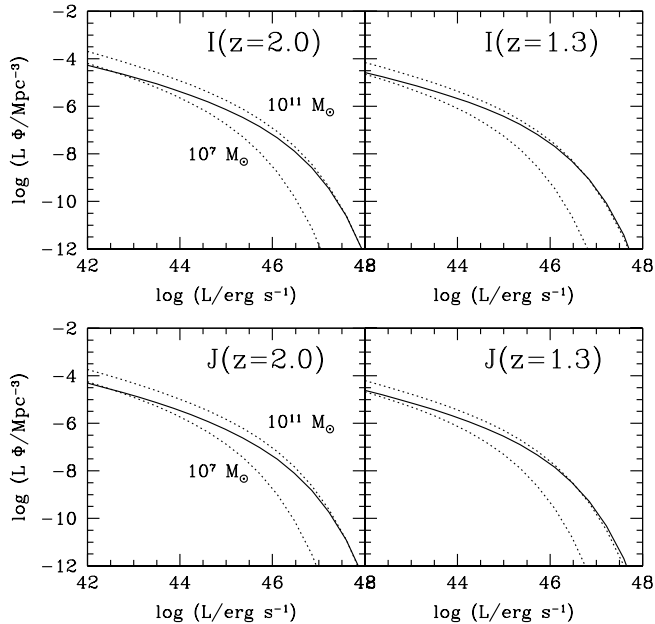}
\caption{QSO LF at I- and J-band at intermediate redshift. 
Dotted lines represents the case that QSOs have the same spectrum
as that for black hole mass of $M_{\rm BH} \simeq 10^7 \rm{and} 
10^{11} M_{\odot}$ at $\dot{M} = 5.0 \times L_{\rm Edd}/c^2$. 
Here, we assume the initial Eddington ratio to be equal to that
at B-band, $\ell=\ell_{\rm B}$.}
\label{levo}
\end{figure}
%%%%%%%%%%%%%%%%%%%%%%%

\section{QSO Luminosity Functions at $z \sim 2$}

Finally, we present the calculated LFs at intermediate redshifts.
Here, we only display the cases with Model A, 
since Model B produces similar results except for the
values of $t_{\rm Q}$ (see figures 3 and 4). 
Model parameters are $t_{\rm Q}= 4.0 \times 10^7$yr 
and $C =4.0 \times 10^{8}$,
which are the same as those in figure 3 ($z \sim 3.0$). 
Then, by setting $\ell$ to be a fitting parameter, 
we can well reproduce the observational QSO
LF at $z=2.0$ and $z=1.3$, as are demonstrated in figure 5.
Here, we let $\ell_{\rm x}$ and $\ell_{\rm B}$ be 
the initial Eddington ratios ($\ell$) in X-ray and B-band, respectively.
The best fit models give $(\ell_{\rm x},\ell_{\rm B}) = (0.24, 0.66)$
at $z=2.0$, 
and $(\ell_{\rm x},\ell_{\rm B}) = (0.10, 0.25)$
at $z=1.3$.
Infrared LFs are calculated by using the same $\ell$ as those of
B-bands and are displayed in figure 6.

Similar to figures 3 and 4, mass dependence of LFs at NIR bands is evident.
The more luminous quasars are, the more massive become central
black holes.
As mentioned before, we have used 
the formation rate of dark halo by Kitayama \& Suto~(1996),
instead the time derivatives of PS mass function. 
The distinction between the two is considerable in these redshifts: 
For example, in X-ray, 
PS mass function under-estimates the number densities
of QSOs at $L_{\rm x} \leq 10^{44} \rm{erg s}^{-1}$ by a factor
of 10 when we choose $\ell$ to fit LFs at higher $L_{\rm x}$ side.
This is because on the lower $L_{\rm x}$ side, negative
destruction rate domiates the time derivative of PS mass function.

%%% Fig. 7 %%%%%%%%%%
\begin{figure}[t]
 \epsfxsize\columnwidth \epsfbox{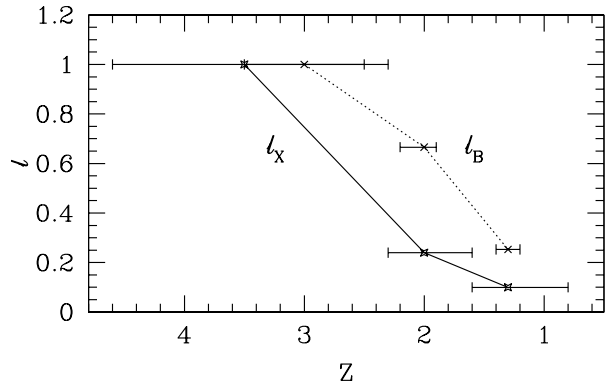}
\caption{
The time evolution of the best fitting initial Eddington ratios,
$\ell_{\rm B}$(dotted line) and $\ell_{\rm x}$(solid line).
%Solid line connects $\ell_{\rm x}$ and dotted line connects $\ell_{\rm B}$.
}
\label{levo}
\end{figure}
%%%%%%%%%%%%%%%%%%%%%%%

Figure 7 plots the time evolution of the initial Eddington ratios. 
Our model requires
$\ell_{\rm B}$ being about twice as much as $\ell_{\rm x}$ at both of
$z=2.0$ and $z=1.3$, which contrasts the cases at $z \sim 3$,
in which the same Eddington ratios, $\ell_{\rm B}=\ell_{\rm x} = 1$, 
can give reasonable fit to the observations.
This implies that for lower mass-flow rates at relatively low redshifts
X-ray radiation tends to be suppressed in the observations with respect to 
our modeled value for the same optical radiation.
In other words,
our model spectra may over-predict X-ray fluxes at $\ell \ll 1$,
if the discrepancy in $\ell_{\rm B}$ and $\ell_{\rm x}$ is real.
However, it is too early to take this difference seriously, since
the observational selection criteria of QSOs are somewhat
different between X-ray and B-band LFs.  Especially, error bars
in X-ray are comparatively large, since smaller number of QSOs were used at
X-ray LFs. We should await future intensive observations to be performed
(see the next section for discussion).

Several issues still remain to be solved.
First, the assumption of $\ell = 1.0$ at $z\sim 3.0$ may not be obvious.
This is a naive expectation, and high redshift QSO survey
may point that $\ell < 1.0$ even at $z \sim 3$.
Next, we have not considered the time evolution of
$t_{\rm Q}$ and $\epsilon$. 
We also did not consider depletion of fuel mass
(Kauffman, Haehnelt 2000; Cavaliere, Vittorini 2000).
These evolution may explain decreasing
LFs at $z \leq 2$ without the evolution of $\ell$.   
  
\section{Discussion}

We construct a simple model for LFs based on the previous models
by HL98 and HNR98 but newly adopting the realistic disk spectra
which have $M_{\rm BH}$ and $\dot M$ dependence.
Our models for LFs can give reasonable fit to the
observed LFs both in B-band and X-rays for redshift $z\sim 3$.
We also show that these LFs are not sensitive to
black-hole mass distribution in QSOs, but instead LFs at NIR,
which will be available in the near future by IR QSO surveys,
should contain information as to $M_{\rm BH}$.

Our model could also reproduce the shape of LFs at 
a variety of redshifts less than $z=3.0$,
however, we need to assume sub-Eddington initial luminosity
of black holes at their formation epoch.
This is consistent as the results obtained by Haiman \& Menou (2000)
who considered the B-band LFs only.  From 
the semi-analytical approach for galaxy/QSO formation, 
Kauffmann \& Haehnelt (2000) suggested that the sharp decline in QSO
number densities from $z \sim 2$ to $z \sim 0$ results from a 
combination of three factors: (i) a decrease in the galaxy-galaxy
merging rate, (ii) a consumption of gas due to star formation
(corresponding to decreasing $\ell$), 
and (iii) an adopted assumption of increasing accretion timescale 
(i.e., decreasing $\ell$ and increasing $t_{\rm Q}$).

Our model predicts $\ell_{\rm x} < \ell_{\rm B}$
by a factor of $\sim 2.5$.  This may indicate that our model
spectra overestimate X-ray flux at luminosity $L \ll L_{\rm Edd}$.
In other words, we need to suppress X-ray flux compared with B-band one
in the model spectra at low luminosities.
%If we try to reproduce LF at lower redshift with $\ell_{\rm x}=\ell_{\rm B}$,
%conversely,
%we will need relatively more X-ray flux as redshift increases. 
This may require significant improvements of the model spectra 
adopted in the present study.
Cosmological evolution of QSO spectrum requires further research.
We also note that this may
take part in the discrepancy of observed comoving density of QSOs at $z>2.5$
between B-band and X-rays.

If QSO accretion-disks turn into the ADAF regime at low redshift due to
low accretion rate (e.g., Yi 1996; Haiman \& Menou 2000),
optical radiation should be much suppressed compared with X-ray radiation;
the broad-band spectrum of ADAF is generally harder than the
present disk-corona spectra, on the contrary to our results.
Intrinsic dust extinction at (rest) optical/UV leads to
the reduction in observed B-band luminosity, 
which again produces the opposite effect. 
%to the fitted $\ell_{\rm x}$ and $\ell_{\rm B}$.
Rather, our result may suggest a drastic shrinkage of the X-ray emitting
region (probably accretion disk corona) at low luminosities.
To probe this, we need more
studies concerning the conditions creating and/or
sustaining hot coronal flow above cool accretion disk.

In the present study, we calculated the formation rates of dark halos by
the formalism based on PS mass function (PS halo),
assuming each PS halo harbors only one black hole.
Generally speaking, however, 
PS halos are not precisely equal to the galactic halos.
For example, the mass scale of $M_{\rm halo} \sim 10^{14} M_{\odot}$
corresponds to a galaxy cluster and, hence, it is not correct to
assume that such a huge PS halo contains only one black hole
(this is known as the over-merging problem). 
Thus, we should be careful with the maximum halo mass $M_{\rm halo}$ 
used in the calculation of QSO LFs so as not to exceed a certain limit.
At $z = 3.0$, the halos which mostly contribute to LFs at the bright end
(for B-band, $L_{\rm B} \sim 10^{13.5} L_{\rm B,sun}$) have 
$M_{\rm halo} \sim 10^{13} M_{\odot}$, but at $z = 1.3$
the maximum halo mass attains $10^{14} M_{\odot}$. 
Therefore,
our calculated LFs suffer over-merging problems
only at higher luminosity side at lower redshift. But to
evaluate $\ell$, we can conclude that this effect is small.
  
Throughout the present study, we assume that each dark halo
contains only one massive black hole whose mass is scaled with
the host dark-halo mass.  It is interesting to note recent discovery
of multiple, intermediate-mass black holes being created in 
post-starburst regions in M82 (Matsumoto \& Tsuru 1999; 
Matsushita et al. 2000).
This is not consistent with our assumption, but it may be that
such intermediate-mass black holes will eventually merge into one big hole
within timescales much shorter than the Hubble time. 

\par
\vspace{1pc} \par
The authors would like to express their thanks to Dr T.\ Miyaji for
providing the electric data of the ROSAT Luminosity Functions.

This work was supported in part 
by Research Fellowships of the Japan Society for the
Promotion of Science for Young Scientists (4616, TK).
% We would like to thank an anonymous referee for valuable comments.

\section*{References}
\small
\re Akiyama et al. 2000, ApJ 532, 700A
\re Abramowicz M., Chen X., Kato S., Lasota J.-P., Regev O. 1995,
    ApJ 438, L37
\re Bertschinger E. 1985, ApJS 58, 39
\re Boller Th., Brandt W.N., Fink H.H. 1996, A\&A 305, 53
%\re Boller Th., Brandt W.N., Leighly K.M., Ward M.J., ed. 2000
%    Proceedings of the Workshop on Observational and Theoretical Progress
%    in the Study of Narrow-Line Seyfert 1 Galaxies, New Astron Rev. 44
\re Boyle B.J., Shanks T., Peterson B.A. 1988, MNRAS 235, 935
\re Boyle B.J., Georgantopoulos,I., Blair,A.J., Stewart,G.C., 
    Griffiths,R.E., Shanks,T., Gunn,K.F., \& Almaini,O. 1998,
    MNRAS 296, 1
\re Brandt W.N., Mathur S., Elvis M. 1997, MNRAS 285, L25
\re Brandt, W. N., et al. 1997, MNRAS 290, 617
\re Cavaliere A., Szalay A.S. 1986, ApJ 311, 589
\re Cavaliere A., Vittorini V. 1998, in ASP Conf. Ser. 146,
    The Young Universe: Galaxy Formation and Evolution at Intermediate
    and High  Redshift, ed. S. D'Odorico, A. Fontana, \& E. Giallongo
    (ASP: San Francisco), p26
\re Cavaliere A., Vittorini V. 2000, ApJ 543, 599
\re Connolly A. Szalay A.S., Dickinson M., Subbarao M.U., Brunner R.J.
    1997, ApJ 486, L11
\re de Bernardis P. et al. 2000, Nature 404, 955
\re Efstathiou G., Rees M.J. 1988, MNRAS 230, 5p
\re Ferrarese L., Merritt D. 2000, ApJ 539, L9
\re Elvis M. et al. 1994, ApJS 95, 1
\re Fontana A. et al. 1998, AJ 115, 1225
\re Franceschini A., Hasinger G., Miyaji T., Malquoli D. 1999
    MNRAS 310, L5
%\re Frank J., King A.R., Raine D.J. 1992, 
%   Accretion Power in Astrophysics (Cambridge Univ. Press, Cambridge).
\re Fukugita et al. 1998, ApJ 503, 518
\re Gebhardt et al. 2000, ApJ 539, L13
\re Grazebrook K. et al. 1999, MNRAS 306, 843
\re Haardt F., Maraschi L. 1991, ApJ 380, L51
\re Haehnelt M.G., Natarajan P., Rees M.J. 1998, MNRAS 300, 817 (HNR98)
\re Haehnelt M.G., Rees M.J. 1993, MNRAS 263, 168
\re Haiman Z., Loeb A. 1998, ApJ 503, 505 (HL98)
\re Haiman Z., Loeb A. 1999, ApJ 521, L9 (HL99)
\re Haiman Z., Menou K. 2000, ApJ 531, 42
\re Ichimaru S., 1977, ApJ 214, 840
\re Kauffman G., Haehnelt M.G. 2000, MNRAS 311, 576
\re Kawaguchi T., Shimura T., Mineshige S. 2001, ApJ 546, 966
\re Kitayama T., Suto Y. 1996, MNRAS 280, 638
%\re Kitayama T., Suto Y. 1996b ApJ 469, 480
\re Kobayashi Y., Sato S., Yamashita T., Shiba H.\ \& Takami H.\ 1993,
    ApJ 404, 94
\re Kormendy J., Richstone D. 1995, ARA\&A 33, 581
\re Lacey C.G., Cole S. 1993, MNRAS 262, 627
\re Laor A. 1998, ApJ 505, L83
\re Laor A., Fiore F., Elvis M., Wikes B. J., McDowell J. C. 1997,
    ApJ 477, 93
\re Macleod et al. 1999, ApJ 511, L67 
\re Madau P. et al. 1996, MNRAS 283, 1388
\re Magorrian J. et al. 1998, AJ 115, 2285 (MG98)
\re Manmoto T., Kusunose M., Mineshige S. 1997, ApJ 489, 791
\re Matsumoto H., Tsuru T. G. 1999, PASJ 51, 321
\re Matsushita S. et al. 2000, ApJ 545, L107
\re Mcleod et al. 1999, ApJ 511, L67
\re Mineshige S., Kawaguchi T., Takeuchi M., Hayashida K. 2000,
    PASJ 52, 499
\re Miyaji T., Hasinger G., Schmidt M. 1998,
    in Proc. Highlights in X-Ray Astronomy (astro-ph/9809398)
\re Miyaji T., Hasinger G., Schmidt M. 2000,
     A\&A 353, 25
\re Narayan R., Yi I. 1994, ApJ 482, L13
\re Neugebauer G.\ et al.\ 1987, ApJS, 63, 615
\re Nulson P.F.J., Fabian A.C 2000, MNRAS 311, 346
\re Osmer P.S. 1982, ApJ 253, 28
\re Otani C., Kii T., Miya K. 1996
    in R\"ontgenstrahlung from the Universe
    (MPE Report 263), ed H.U. Zimmermann, J.E. Tr\"umper, H. Yorke
    (MPE Press, Garching) p491
\re Pei Y.C. 1995, ApJ 438, 623
\re Polletta, M. \& Courvoisier, T.J.-L. 1999, A\&A, 350. 765
\re Pounds K.A., Done C., Osborne J. 1996, MNRAS 277, L5
\re Press W.H., Schechter P.L. 1974, ApJ 181, 425
\re Radovich M., Klaas U., Acosta-Pulido J. \& Lemke D.\ 1999, AA, 348, 705
\re Rees M.J. 1984 ARA\&A 22, 471
\re Rees M.J., Blandford R.D., Begelman M.C., Phinney S. 1982,
    Nature 295, 17
\re Sasaki S. 1994, PASJ 46, 427
%\re Schmidt 1989
\re Schmidt M., Green R.F. 1983, ApJ 269, 352
\re Shakura N.I., Sunyaev R.A. 1973, A\&A 24, 337
\re Shaver P.A. et al. 1996, Nature, 384, 439
\re Shaver P.A., Hook I.M., Jackson C.A., Wall J.V., Kellermann K.I.
    1999 in Highly Redshifted Radio Lines, eds. C. Carilli et all.
    (PASP: San Francisco), p163
\re Steidel et al. 1999, ApJ 519, 1
\re Stengler-Larrea, E. A. et al. 1995, ApJ 444, 64
\re Telesco C.M., Becklin E.E., Wynn-Williams C.G. \& Harper D.A.\ 1984 
 ApJ 282, 427
%\re Tsuru et al. 1996
\re Veilleux, S., Sanders, D. B., \& Kim, D.-C. 1999, ApJ 522, 139
\re Yi I. 1996, ApJ 473, 645
\re Zheng W., Kriss G.A., Telfer R.C., Grimes J.P.,
    Davidsen A.F. 1997, ApJ 475, 469
\end{document}